\documentclass[aps,amsmath,amssymb,12pt]{revtex4}
\usepackage{graphicx}
\usepackage{bm}

\newcommand{\be}{\begin{equation}}
\newcommand{\ee}{\end{equation}}
\newcommand{\bea}{\begin{eqnarray}}
\newcommand{\eea}{\end{eqnarray}}
\newcommand{\hf}{\frac12}
\newcommand{\nn}{\nonumber\\}

\newcommand\eqn[1]{Eq.\,(\ref{#1})}

\newcommand\fig[1]{Fig.\,{\ref{#1}}}
\newcommand\sect[1]{Sect.\,{\ref{#1}}}

\newcommand{\beq}{\begin{equation}}
\newcommand{\eeq}{\end{equation}}

\def\tu{{\tilde u}}

\def\ord#1{{\cal O}(#1)}

\def\mr#1{{\mathrm{#1}}}

\def\Tr{{\mathrm{Tr}}}
\def\ord#1{{\cal O}\left(#1\right)}
\def\mr#1{{\mathrm{#1}}}

\def\fd#1#2{\frac{\delta#1}{\delta#2}}

\def\beq{\begin{equation}}
\def\eeq{\end{equation}}
\def\dk{\Delta k}

\def\dk{{\Delta k}}

\def\oh{\frac1\hbar}

\def\t{\tilde}
\def\L{\Lambda}

\begin{document}
\title{Renormalization of the bilocal sine-Gordon model}

\author{I. Steib, S. Nagy}
\affiliation{Department of Theoretical Physics, University of Debrecen,
P.O. Box 5, H-4010 Debrecen, Hungary}
\date{\today}

\begin{abstract}
The functional renormalization group treatment is presented for the two-dimensional sine-Gordon model by including a bilocal term in the potential, which contributes to the flow at tree level. It is shown that the flow of the bilocal term can substitute the evolution of the wave function renormalization constant, and then the Kosterlitz-Thouless type phase transition can be recovered.
\end{abstract}

\maketitle

\section{Introduction}\label{sect:intro}

The quantum field theoretical models suffer from ultraviolet (UV) divergences, and we should remove the infinities by regularizations. They imply that we should consider an effective dynamics of the models, where certain degrees of freedom or modes of the physical system are not followed, they do not belong to the system anymore, they are pushed to the environment. However, if we integrate out certain degrees of freedom, then the remaining effective model of the system modes becomes nonlocal \cite{Polonyi:2017eaz}. Earlier works investigated nonlocal field theories at perturbative level \cite{Yukawa:1950eq,Yukawa:1950er}. Nevertheless, the theories including nonlocal interactions have to face with the violation of causality \cite{Efimov:1967pjn,Alebastrov:1973np,Tomboulis:2015gfa}, instability problems \cite{Eliezer:1988rr,Eliezer:1989cr}, or the spoiling of gauge invariance \cite{Moffat:1990jj,Evens:1990wf,Kleppe:1991rv}.

The introduction of a momentum cutoff, as a simple regularization, also leads to nonlocal interactions. The same problem takes place, when we use the functional renormalization group (RG) method \cite{Wetterich:1992yh,Berges:2000ew,Polonyi:2001se,Pawlowski:2005xe,Gies:2006wv}. Recently it has been pointed out, that RG blocking step introduces nonlocal contributions to the evolution \cite{Polonyi:2006yw,Nagy:2015rva,Nagy:2018fdh}. During the blocking some ultraviolet (UV) modes in the momentum shell are integrated out, transforming system degrees of freedom into the environmental ones. The nonlocality has not been followed during the traditional the RG method, because we usually use local actions and do not let the nonlocal terms to evolve. However there is a nontrivial saddle point for the integrated UV modes, which introduces a nonlocal term, moreover it contributes  to the RG evolution at tree level.

The nonlocality can cause significant changes in the properties of the investigated models. Old fixed points of the local models can vanish, new fixed points can appear, which can even introduce new phases in the model. This is the case for the 3-dimensional (3d) $\phi^4$ theory, which is one of the most widely investigated quantum field theoretical model. The nonlocality can be introduced in the simplest way by introducing a bilocal potential into the action. Due to bilocality the Wilson-Fisher fixed point disappears in the model, and a new hyperbolic type fixed point appears close to the separatrix belonging to the first order phase transition. Furthermore, the bilocality introduces new relevant interactions \cite{Nagy:2018fdh}. These results clearly show the significance of the nonlocality, and suggests that we should repeat the RG investigation including the evolution of the bilocal potential for every model.

The nonlocality can be used with the sharp cutoff, where the saddle point can be obtained analytically from the linearized equations of motion for the UV modes. When we have smooth cutoff, then the linearization cannot be used, therefore the saddle point can be determined only numerically, which makes the evolution equations practically impossible to follow. These reasons make impossible to include the bilocal terms for the smooth cutoff. However the smoothness enables us to go beyond the local potential approximation (LPA) and we can calculate the evolution of the wave function renormalization and further terms in the gradient expansion. The gradient expansion mimics the nonlocality in the sense, that it gives correction to the constant fields used in LPA. As the we include higher and higher orders in the derivatives we can smear the locality of the interaction in its infinitesimal environment. The bilocal term goes further, where arbitrary nonlocal interactions can be treated. In this sense the RG method using sharp cutoff and including a bilocal potential serves a more general treatment for the models than the smooth cutoff RG technique with arbitrary order of the gradient expansion. Consequently it is worth reinvestigating the models which were treated earlier by the usual RG technique using gradient expansions. The example of the 3d $\phi^4$ model shows that we should expect significant changes in every aspect of the model.

The 2-dimensional (2d) sine-Gordon (SG) model in Euclidean spacetime is one of the most important scalar model, since it is widely used in almost every area of modern physics  \cite{Coleman:1974bu,Amit:1979ab,Nandori:1999vi,Nagy:2006pq,Pangon:2009wk}. Furthermore, some common aspects of the model with the non-Abelian gauge theories suggests that the confinement mechanism can be understood in a simpler framework \cite{Nandori:1999vi,Nagy:2006pq,Pangon:2009wk}. Another interesting point is that the model exhibits a Kosterlitz-Thouless (KT) type infinite order phase transition \cite{Berezinskii:1971,Kosterlitz:1973xp}. Interestingly the RG treatment shows significant difference if we compare the results obtained by LPA and including the lowest order term in the gradient expansion. In LPA we can find two phases of the SG model, separated by a critical value of the wave number parameter $\beta_c^2=8\pi$, the Coleman point \cite{Coleman:1974bu}, and the phase space contains straight vertical lines. If we let wave function renormalization constant to evolve, then, instead of the straight lines we get hyperbolas in the phase space. Furthermore, the essential scaling of the correlation length $\xi$ can be recovered.

However, there is a very important result in the SG model, which cannot be obtained in the traditional RG method, namely the special role of the parameter $\beta^2=4\pi$. The SG model represents the bosonized version of the 2d massive noninteracting Thirring model for $\beta^2=4\pi$ \cite{Mandelstam:1975hb}. The bosonization technique can change the fermionic degrees of freedom into bosonic ones, and in $d=2$ the transformation introduces local bosonic fields.  During the RG flow of the SG model  the trajectories tend to the IR fixed point in the broken symmetric phase, independently on the initial value of $\beta$, and show nothing speacial for $\beta^2=4\pi$ \cite{Nagy:2009pj}. Nevertheless a less precise analysis, which uses LPA can exhibit some difference between the scalings below and above $4\pi$. When $4\pi<\beta^2<8\pi$, then there is a nontrivial IR fixed point for finite value of $\t u$, and for $\beta^2<4\pi$ the evolution runs into singularity. We should admit that the flow equation approach of the RG technique can account for the special value of $\beta^2=4\pi$, since it gives such an evolution, where the IR fixed point is situated at $\beta^2=4\pi$ in the broken symmetric phase \cite{Kehrein:1999nx}. Unfortunately, the value $\beta^2=4\pi$ is not prominent in the standard RG calculations, and we should complete the RG treatment of the 2d SG model in order to eliminate this shortcoming. 

The significant difference between the phase structures of the SG model in LPA and with gradient expansion suggests the necessity of the nonlocal RG treatment. We expect that the evolution of a bilocal term in the SG model can give us the essential scaling of the correlation length $\xi$. Our goal is to determine the evolution of the SG model with a bilocal potential and to show, that the model recovers the standard KT type phase structure, and the special role of $\beta^2=4\pi$. The advantage of our treatment relies in its simplicity. It is extremely difficult to determine the contributions of the gradient expansion both technically and numerically. However, the calculation of the RG evolution for the bilocal potential is shamelessly trivial, and can be performed analytically. We argue that the RG method, using LPA + bilocal potential (in the tree level), should replace the traditional RG technique with the gradient expansion. This suggestion is nicely demonstrated in our the RG treatment for the SG model.

The paper is organized as follows. In \sect{sect:locsg} we introduce the RG method and its results for the SG model. In \sect{sect:sp} the role of the nontrivial saddle point is discussed. The results for the tree level evolution is presented in \sect{sect:sgtree}, and ground state is calculated in \sect{sect:gs}.  Finally, in \sect{sect:sum} the conclusions are drawn up.

\section{The local sine-Gordon model}\label{sect:locsg}

The 2d SG model in Euclidean spacetime, is defined by the Wilsonian action
\beq\label{sgact}
S =\int_x \left[\hf (\partial_{\mu}\phi)^2 +U_k(\phi_x)\right],
\eeq
with $\int_x = \int x^d x$. The periodic local potential has the form
\be\label{Upot}
U_k(\phi_x) = u \cos(\beta\phi).
\ee
It has a $Z_2$ symmetry $\phi_x\to-\phi_x$ and it is periodic in the internal space, $\phi_x\to\phi_x+2\pi/\beta$, with the wave number parameter $\beta$. The periodicity is kept during the RG procedure.

The SG model belongs to the same universality class as the 2d Coulomb gas and the 2d XY spin model \cite{Nandori:2000rx}. It can account for the KT type phase transition of vortices in a thin superfluid film. The SG model has two phases, separated by the Coleman point $\beta^2_c=8\pi$. Above the Coleman point we have a symmetric phase, while below $\beta_c^2$ the broken symmetric phase can be found. In the XY spin model the phases can be identified by different layouts of the vortices. We can find dissociated vortices and anti-vortices in the symmetric phase, and in the broken symmetric phase the vortices and antivortices form bound pairs \cite{Huang:1990via}. The vortices can be considered as the elementary nonlocal excitations formed by revolving spins around a point. The vortices and its nonlocal nature seem to play a crucial role to describe the phase structure and the topological, KT type phase transition in the XY model. Although the connection between the SG and the XY model is indirect, we expect that the phase structure of the SG model cannot be understood well without taking into account nonlocal interactions.

\subsection{Local potential approximation}\label{ssect:evol}

We use the Wegner-Houghton (WH) equation \cite{Wegner:1972ih}, where the evolution of the Wilsonian action is followed by lowering the gliding cutoff $k$. We split the field variable into the sum $\phi\to\phi+\varphi$, where $\phi$ is identified as the system variables, the IR modes, which is nonvanishing for momenta $0<|p|<k-\dk$, while $\varphi$ belongs to the environmental field variables, the UV components, with momenta $k-\dk<|p|<k$. The sharp cutoff, used in the WH equation, can separate the system and the environmental variables uniquely. The smooth cutoff mixes the UV and the IR modes during the elimination, which can cause difficulties if we would like to identify the entanglement between the system and the environmental modes \cite{Nagy:2015rva}. The blocked action is given by
\be\label{blocking}
e^{-S_{k-\dk}[\phi]}=\int D[\varphi]e^{-S_k[\phi+\varphi]}\approx e^{-\oh S_k[\phi+\varphi[\phi]]-\hf\Tr\ln S''_k},
\ee
where $\varphi[\phi]$ is the saddle point (if there is any), and the notation  $^\prime=\partial/\partial\varphi$ is used. We start from a blocked action $S_k$ and decrease the value of the gliding cutoff scale $k$ by an infinitesimal step $\dk$. During one blocking step we integrate out the modes in the momentum shell $[k-\dk,k]$ transforming the system degrees of freedom to the environment. 
The action contains the usual kinetic term and a local potential, $S=S_0+S_1$, i.e.
\be
S_0[\phi]=-\hf\int_x\phi_x\Box\phi_x,
\ee
and
\be
S_1[\phi]=\int_x U(\phi_x).
\ee
\eqn{blocking} leads to the WH equation in LPA
\be\label{wh}
\dot U = -\alpha_d k^d \ln(k^2+U''),
\ee
where the dot stands for $k\partial_k$, and the constants $\alpha_d=1/2^d\pi^{d/2}\Gamma[d/2]$ is introduced. We can linearize the the RG equation in the potential and obtain
\be\label{Ulin}
\dot U = \frac1{4\pi} U'',
\ee
when $d=2$. Inserting the periodic potential in \eqn{Upot} into \eqn{Ulin} we get the evolution equation for the coupling
\be\label{ulinsol}
\dot{\t u} = -2\t u+\frac{\beta^2}{4\pi}\t u.
\ee
Its analytical solution is given by
\be\label{linsolu}
\t u = \t u(\Lambda) k^{-2+\beta^2/4\pi},
\ee
where the scale $k$ runs from $\Lambda$ towards zero, $\t u(\Lambda)$ gives the initial value of the coupling, and the symbol $\t~$ stands for the dimensionless quantities, e.g. $\t u = u/k^2$. The cutoff $\Lambda$ represents the upper momentum scale, above which the modes assumed to be unimportant. From the exponent of the \eqn{linsolu} we can identify the Coleman point $\beta^2_c=8\pi$, where the coupling $\t u$ is marginal and does not evolve \cite{Nandori:1999vi,Nagy:2006pq,Pangon:2009wk}. Above the Coleman point, in the symmetric phase, the coupling $\t u$ tends to zero.  In the symmetric phase the coupling $\t u$ is irrelevant, therefore this phase is nonrenormalizable in perturbative sense. Below the Coleman point the coupling tends to infinity, it is relevant, and the trajectories correspond to the broken symmetric phase.

The exact evolution equation in \eqn{Ulin} can also be obtained analytically. Interestingly the position of the Coleman point does not change \cite{Nagy:2009pj}. There, the phase space can be completed by an IR and a UV non-Gaussian fixed points \cite{Kovacs:2014fwa}, althought there regimes are far from the physically interesting Coleman point, furthermore, the trajectories of the broken symmetric phase run into singularity, i.e. $\t u \to -1$ in the IR limit. The fixed point remains unchanged when we take into account the effects of the upper harmonics \cite{Nagy:2006pq}, nevertheless they can give the correct form of the negative parabola for the IR effective potential in the broken symmetric phase.

We note that the Coleman point can be obtained in the leading order solution of the RG equation. Its reason comes from the fact, that the tree level scaling of the couplings changes. The leading order scaling is determined by the canonical dimension of the couplings. From the linearized RG equation in \eqn{Ulin} we can get corrections to the tree level scaling if $U''\sim U$. This is the case for the SG model, therefore the canonical dimension $\t u \sim k^{-2}$ should be completed by the exponent $\beta^2/4\pi$, which leads to the Coleman point. In polynomial models the term $U''$ contributes to couplings of the higher order interactions, therefore the leading order scalings can give only qualitative results, the loop corrections change the phase structure and the position of the fixed point of the model significantly.

\subsection{Wave function renormalization}

In order to incorporate the evolution of the wave function renormalization constant $z$, we should use the RG technique for the effective action given by the Wetterich equation \cite{Wetterich:1992yh,Morris:1993qb,Berges:2000ew}. It gives the scale dependence of the effective action
\beq\label{feveq}
\dot \Gamma=\hf\mr{Tr}\frac{\dot R}{\Gamma''+R},
\eeq
where the trace Tr denotes the integration over the momenta. A power-law type regulator function is used,
\beq
R = p^2\left(\frac{k^2}{p^2}\right)^b,
\eeq
with the parameter  $b\ge 1$. We assume that the effective action in \eqn{feveq} has the same functional form as the Wilsonian action, i.e.
\beq\label{eaans}
\Gamma = \int_x\left[\frac{z}2 (\partial_\mu\varphi_x)^2+U_k(\phi_x)\right],
\eeq
with similar local potential as in \eqn{Upot}, and the field-independent wave-function renormalization $z$. After a simple rescaling of the field variable, we obtain $z=1/\beta^2$ for the initial value. \eqn{feveq} leads to the evolution equations \cite{Nagy:2006ue}
\bea\label{ea_v}
\dot U &=& \hf\int_p{\cal D}_k\dot R,\\
\label{ea_wf}
\dot z &=&
{\cal P}_0 U'''^2_k\int_p{\cal D}_k^2\dot R\left(
\frac{\partial^2{\cal D}_k}{\partial p^2\partial p^2}p^2
+\frac{\partial{\cal D}_k}{\partial p^2}
\right),
\eea
where ${\cal D}_k=1/(zp^2+R+U'')$, and ${\cal P}_0=(2\pi)^{-1}\int_0^{2\pi} d\phi$  projects onto the field-independent subspace. Again, we keep only the leading order terms in $U$ and consider the fundamental mode in the periodic potential. After performing the 2d momentum integrals, we obtain that
\bea
\label{ulin}
\dot{\t u} &=& -2 \t u+\frac1{4\pi z}\tu, \\
\label{zlin}
\dot z &=& -\frac{\tu^2}{z^{2-2/b}}c_b,
\eea
with the constant $c_b = b\Gamma(3-2/b)\Gamma(1+1/b)/48\pi$ \cite{Nagy:2009pj}. The phase space trajectories become hyperbolas on the $(z,\t u)$ plane. The KT type phase transition is characterized by the exponential dependence of the correlation length $\xi$ on the inverse of the square-root of the reduced temperature. In this model we obtain that $t\sim\t u^{* 2}$, where $\t u^*$ is the value of the coupling at the turning point of the trajectories in the broken symmetric phase.

We note that in the sharp cutoff limit the r.h.s. of \eqn{zlin} diverges as $b\to \infty$. It signals that we need a smooth cutoff to determine the evolution of $z$, we have no wave function 
renormalization for the sharp cutoff. We have an opposite situation for the bilocal potential, it can be easily adopted to the WH equation with sharp cutoff, and can hardly be used for the Wetterich equation with smooth cutoff. Furthermore the bilocal potential serves a more general and simpler treatment of nonlocality than the gradient expansion, as we demonstrate it in the next section.

\section{Saddle point}\label{sect:sp}

In the spirit of the WH equation the environmental modes in the $d$-dimensional sphere shell are eliminated in infinitesimal RG blocking steps. We can find a saddle point in the eliminated momentum shell by solving the equation of motion for the UV mode $\varphi$ for a given IR mode $\phi$ \cite{Nagy:2015rva,Nagy:2018fdh}. Due to the sharp cutoff it is enough to find the saddle point of the linearized equations of motion, since for any $f_x$
\be\label{vphidk}
\int_xf_x\varphi^n_x=\int_qf_q\prod_{j=1}^n\int_{k-\dk<|q_j|<k}\varphi_{q_j}\delta_{q,-\sum_jq_j}
=\ord{\dk^n}.
\ee
The result shows that the $n$th order expression in $\varphi_k$ is proprtional to $\dk^n$, which is negligible, only the term linear in $\varphi_k$ contributes to the solution. The nontrivial saddle point can enlightened by considering the equation of motion with the momenta of the field variables. For simplicity  let's first formulate the equation of motion for the $\phi^4$ theory with the Euclidean action
\be
S_k[\phi]=-\hf\int_x\phi_x\Box\phi_x+\int_x\left(\frac{m^2}2\phi^2_x+\frac{g}{4!}\phi^4\right),
\ee
where the field variables are redefined again as the sum of the system and the environmental modes, $\phi\to \phi+\varphi$. Taking $-\Box+m^2=D_0^{-1}$ the equation of motion is
 \be\label{treephi4}
 D_0^{-1}(\phi_p+\varphi_k)=-\frac{1}{6}g(\phi_{p}^3+3\phi_{p}^2 \varphi_{k}+3\phi_{p} \varphi_{k}^2+\varphi_{k}^3),
 \ee
where the momenta of the UV modes $\varphi_k$ are taken from the momentum shell to be eliminated, and the IR modes have momentum from the region $p\in [0,k-\dk]$. The quadratic term connects an IR with an UV fields with the same momenta, which would break the momentum conservation, therefore they do not contribute to the equation. In order to satisfy equation for the momenta on both sides we should choose $p=k/3$ in the absence of the zero modes. The remaining, quartic terms provide the saddle point
\be
\varphi_k = -\frac{1}{6}gD_0\phi_{k/3}^3.
\ee
It shows that the interaction vertex contains a leg belonging to the UV mode, which splits into 3 legs of IR modes. Two vertices constitute a bilocal, six-leg vertex, with an UV propagator in the middle. It implies, that we need at least a 6th order vertex in order to find a nontrivial saddle point contribution. The higher order vertices matter more to the saddle point, since the momentum $k$ of the UV field can be combined from many IR momenta $p$ more easily.

The nontrivial saddle point gives a tree level contribution to the RG evolution, since the change of the action during the blocking can be given as
\be
S_{k-\dk}[\phi]=S_k[\phi+\varphi],
\ee
giving
\be\label{preDS}
\Delta S \approx  - \hf  \varphi S^{(2)}\varphi.
\ee
The change is bilocal, and is usually suppressed in the traditional RG treatment. However it modifies the evolutions at tree level, which is definitely more important than the usual loop contributions. If we would like to follow their evolution, we should introduce a bilocal potential as
\be
S_2[\phi]=\int_{xy} V_{x,y}(\phi_x,\phi_y).
\ee
Thus the blocked action can be written as the sum of the kinetic, local, and bilocal terms, $S=S_0+S_1+S_2$. We assume a translation invariance for the coordinate dependence, $V_{x,y}=V_{x-y}$. In momentum space, after the Fourier transformation the bilocal potential picks up a momentum dependence,
\be
V_q(\phi_1,\phi_2)=\int_xe^{iq(x-y)}V_{x-y}(\phi_1,\phi_2).
\ee
It implies that we have infinitely many couplings due to the continuous value of $q$. If we look for fixed points, then we should consider all the couplings with every value of $q$.

Fortunately the idea of finding the nontrivial saddle point can be easily generalized into any types of bilocal potentials \cite{Nagy:2018fdh}. In general case the equation of motion becomes
\be
\fd{S}{\varphi}=D_0^{-1}\varphi+U'(\phi_x+\varphi_x)+2\int_y\partial_1V_{x-y}(\phi_x+\varphi_x,\phi_y+\varphi_y)=0,
\ee
and after the linearization we obtain
\bea
0&=&\int_y\left[D^{-1}_{0x-y}+\delta_{xy}U''(\phi)+2\delta_{xy}\int_z\partial^2_1V_{x-z}(\phi_x,\phi_z)+2\partial_1\partial_2V_{x-y}(\phi_x,\phi_y)\right]\varphi_y\nn
&&+U'(\phi_x)+2\int_y\partial_1V_{x-y}(\phi_x,\phi_y).
\eea
We introduce the full propagator as
\be
D^{-1}_{xy}=D^{-1}_{x-y}+\delta_{xy}U''(\phi)+2\delta_{xy}\int_z\partial^2_1V_{x-z}(\phi_x,\phi_z)+2\partial_1\partial_2V_{x-y}(\phi_x,\phi_y),
\ee
while the remaining part is denoted by
\be
L_x=U'(\phi_x)+2\int_y\partial_1V_{x-y}(\phi_x,\phi_y).
\ee
By using this notation the nontrivial saddle point can be written in a very simple form of
\be
\varphi=-DL.
\ee
From \eqn{preDS} it gives
\be\label{Dktree}
\Delta S = -\frac{\dk}2\int_{xy}\int_{|p|=k}L_x D_{p}e^{-ip(x-y)} L_y.
\ee
The momentum integration is restricted to the momentum shell to be integrated in an RG blocking step. For simplicity we can introduce the environmental propagator
\be
D_{x-y}^{(k)} = \int_p \delta(|p|-k)D_{p}e^{-ip(x-y)}.
\ee
The infinitesimal change of the action in \eqn{Dktree} contributes to the bilocal potential, and at tree level its rate of change is
\be\label{treeev}
\frac{\Delta V^{tree}_q(\phi_1,\phi_2)}{\dk}=-2D^{(k)}_{q}\left[\partial_2V_{q}(\phi_1,0)+\hf U'(\phi_1)\right]\left[\partial_1V_{q}(0,\phi_2)+\hf U'(\phi_2)\right].
\ee

\section{Tree level evolution of the bilocal sine-Gordon model}\label{sect:sgtree}

\begin{figure}
\includegraphics[width=10cm]{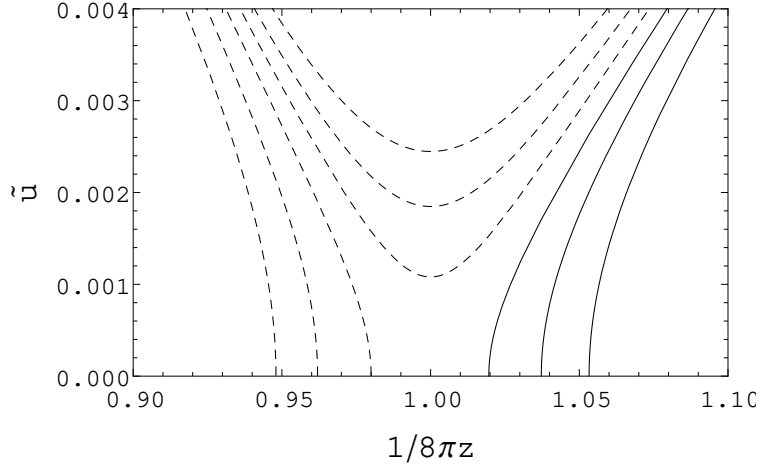}
\caption{The phase structure of the SG model is presented. The solid (dashed) trajectories belong to the symmetric (broken symmetric) phase, respectively.}\label{fig:phase}
\end{figure}

In the previous section a general framework has been given how to calculate the the tree level bilocal contribution to the RG evolution, and now we apply them to the 2d SG model. The Euclidean action contains a massless kinetic term, a local periodic potential in \eqn{Upot}, and a bilocal potential
\be\label{sgpots}
S_2=\int_{xy}v_{x-y}\sin(\beta\phi_{x})\sin(\beta\phi_{y}).
\ee
The zero momentum of the bilocal potential $V_0$ evolves only at loop level. Then, the WH equation reduces to the form in \eqn{wh}, containing only the local potential. Keeping the leading order term in the evolution we obtain back the analytic solution in \eqn{ulinsol} for the local coupling $\t u$. The evolution of the bilocal potential can be obtained from \eqn{treeev}, and it becomes
\be
\Delta{V}=-\frac{2\beta^2\sin(\beta\phi_{1})\sin(\beta\phi_{2})(v-\frac{u}{2})^2}{k^2-\beta^2 u+2\beta^2 v}.
\ee
After the identification $\beta^2\to 1/z_0$ and keeping only the leading order terms, we obtain
\be\label{Dvq}
\Delta{v_{q}}=-\frac{u^2}{2z_0k^2}\omega_{q,k},
\ee
where we turned from coordinate to momentum dependence in $v$, and introduced the square function $w_{q,k}$. The tree level contributions for certain values of the momentum $q$ are independent, so we should only sum up them to get the total IR results. If we neglect the bilocal loop contribution, then only the momentum $q=k$ is nontrivial \cite{Nagy:2018fdh}, and we obtain the relation
\be
v_k=\frac{u^2}{2z_0k^2}\frac{\dk}{k}.
\ee
The inverse full propagator from the Wilsonian action, including the bilocal contribution is
\be
D_{0}^{-1}=z_0k^2-u+2v,
\ee
with the initial parameter $z_0$, coming from the wave number $\beta$. If we compare it with the inverse propagator 
\be
D_{0}^{-1}=zk^2-u,
\ee
with the running the wave function renormalization, we can see, that its evolution can be related by the parameters $z_0$ and $v$ according to
\be
z-z_0=\frac{u^2}{k^2z_0}\dk.
\ee
which gives the following dimensionless evolution equation for the wave function renormalization
\be
\dot z = -\frac{\t u^2}{z}.
\ee
It can be identified with the case of $b=2$ in \eqn{zlin}. We note that this gives the optimized value in the regulator. Our results can give the correct form of the evolution of $z$ obtained by using the gradient expansion of the effective action. There is an analytic solution in the $z,\t u$ plane, we obtain hyperbolas satisfying
\bea\label{uzlin}
\tu^2(z) = 2(z-1/8\pi)^2+\t u(z_0)^{2}.
\eea
The trajectories are shown around the Coleman point in \fig{fig:phase}.
We recovered the usual phase structure of the KT type phase transition. Naturally our treatment is restricted to the close vicinity of the fixed point.

\section{Ground states}\label{sect:gs}

\begin{figure}
\includegraphics[width=10cm]{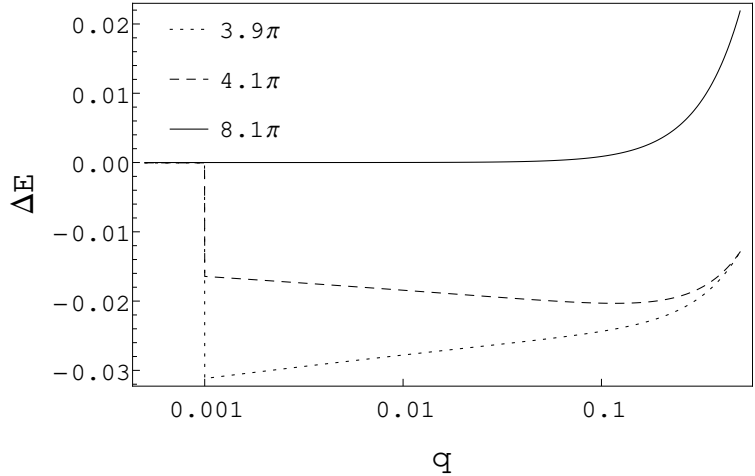}
\caption{The energy difference is presented for different values of $z_0$. The solid line shows the energy for the symmetric phase. The dotted line depicts the energy for $1/z_0=3.9\pi$. It has a nontrivial minimum at $q=k=0.001$. The dashed line corresponds to $1/z_0=4.1\pi$, the nontrivial minimum situated at $q=0.106$.}\label{fig:gs}
\end{figure}

We identify the Euclidean action by the energy of the system at semiclassical level. Its minimum can be identified by the ground state of the system. For simplicity we look for the minimum among periodic field configurations, $\phi_x=\phi\cos(qx^1)$. When the wave number $q$ is zero, then we get the energy for the constant field. The form of the energy difference related to the constant field is
\be
\Delta E = \frac14 q^2 \phi^2-2J_0(\pi)u(\cos\phi-1) + \int_q v_q 4 J_0^2(\pi)\sin\phi_1\sin\phi_2,
\ee
if we take only the leading order Fourier terms. $J_0$ denotes the Bessel function of the first kind, and the integral of $v_q$ is calculated from \eqn{Dvq}. The energy minimum is searched in the $\phi,q$ plane. The results are shown in \fig{fig:gs}.
The energy exhibits different behavior for different values of $z_0$. In the symmetric phase for $1/z_0>8\pi$ we obtain that the minimum can be found at $q=0$, therefore the ground state is trivial. There are certain initial values, where a false minimum appears at $q=\L$, but these values are excluded, since they correspond to the environment after the RG blocking. In the broken symmetric phase we can have a nontrivial ground state. However we can distinguish two regions. It implies that the model has three phases. When $1/z_0<4\pi$ then, although we have a nontrivial minimum, it is always situated at the scale $k$, $q=k$. In the IR limit $q$ tends to zero, and the symmetry of the ground state is restored. In the $4\pi<1/z_0<8\pi$ case there is a finite value of $q$ where the energy has a nontrivial minimum, the ground state is given by a periodic field configuration. The fixed value of the wave number and the periodicity suggests that we have an antiferromagnetic ground state in this phase in the language of the spin models. The finite $q$ implies that the bilocal potential generates an inhomogeneous ground state. At $1/z_0=4\pi$ the curve behaves as in the symmetric phase.

The nontrivial ground state shows, that now we can account for the special value $1/z_0=4\pi$, which corresponds to the value, where the bosonization connects the SG model with the Thirring model \cite{Nandori:2007zs,Nandori:2010ij}. There the bilocal coupling does not evolve, which suggests that the SG model becomes local for $1/z_0=4\pi$.

We can conclude, that the bilocal treatment of the SG model can account for the KT type phase transition around the Coleman point, furthermore it also shows the special role of $1/z_0=4\pi$.

\section{Summary}\label{sect:sum}

The functional renormalization group treatment of the 2-dimensional sine-Gordon model in the framework of the Wegner-Houghton equation is improved by including the evolution of the bilocal term in the action. The RG blocking steps introduce a nontrivial saddle point, which gives a tree level contribution to the RG evolution to the bilocal potential. We showed that the bilocal term can account for  the Kosterlitz-Thouless type essential scaling of the correlation length $\xi$. This result could only be obtained in the framework of the Wetterich equation with smooth cutoff and taking into account the gradient expansion beyond the local potential approximation.

The nonlocal interactions modify the phase structure of the models significantly. This has been demonstrated for the sine-Gordon model. The phases, and the essential scalings obtained by gradient expansion could be recovered by the tree level contribution coming from the bilocal potential. We also recovered the special role of $\beta^2=4\pi$, where the sine-Gordon model is equivalent to the Thirring model.

The loop corrections for the tree level evolution may modify the phase structure further. We note, that the scalings around the Coleman point has physical relevance and many applications. However, as we go away from the linear region of the fixed point, although the flows are well known, they possess less physical relevance.

As we argued, the nonlocality should be taken into account in every field theoretical models. However its role is inevitable, where the elementary excitations themselves are nonlocal. As an example we used the bilocal treatment for the sine-Gordon model which can be related to the planar XY model where the nonlocal vortex interaction can describe the model behavior. It suggests, that some low dimensional models can be investigated more precisely if we include the effects of nonlocality. It may confirmed by the fact, that we should take more and more terms in the gradient expansion to get reliable results as the dimension is lowered from $d=4$ down to $d=2$ or even lower values.

The nonlocality has a central role, when we formulate the renormalization procedure in Minkowski spacetime. The Minkowski propagator contains off-shell and on-shell contributions. The latter is diagonal in the momentum space, therefore cannot be local in the coordinate space. The correct treatment of renormalization in Minkowski spacetime requires the introduction of nonlocal actions. It implies that the we loose the contributions of the on-shell excitations in Euclidean formalism, therefore the renormalization group investigations should be performed in Minkowski spacetime.

\section*{ACKNOWLEDGMENTS}
I. Steib acknowledges the support of the \'UNKP-18-3 New National Excellence Program of the Ministry of Human Capacities. S. Nagy acknowledges financial support from the Hungarian National Research, Development and Innovation Office NKFIH (Grant Nos. K112233, KH126497).

\bibliography{steib}

\begin{thebibliography}{37}
\expandafter\ifx\csname natexlab\endcsname\relax\def\natexlab#1{#1}\fi
\expandafter\ifx\csname bibnamefont\endcsname\relax
  \def\bibnamefont#1{#1}\fi
\expandafter\ifx\csname bibfnamefont\endcsname\relax
  \def\bibfnamefont#1{#1}\fi
\expandafter\ifx\csname citenamefont\endcsname\relax
  \def\citenamefont#1{#1}\fi
\expandafter\ifx\csname url\endcsname\relax
  \def\url#1{\texttt{#1}}\fi
\expandafter\ifx\csname urlprefix\endcsname\relax\def\urlprefix{URL }\fi
\providecommand{\bibinfo}[2]{#2}
\providecommand{\eprint}[2][]{\url{#2}}

\bibitem[{\citenamefont{Polonyi}(2017)}]{Polonyi:2017eaz}
\bibinfo{author}{\bibfnamefont{J.}~\bibnamefont{Polonyi}},
  \bibinfo{journal}{EPL} \textbf{\bibinfo{volume}{120}}, \bibinfo{pages}{40005}
  (\bibinfo{year}{2017}), \eprint{1708.02027}.

\bibitem[{\citenamefont{Yukawa}(1950{\natexlab{a}})}]{Yukawa:1950eq}
\bibinfo{author}{\bibfnamefont{H.}~\bibnamefont{Yukawa}},
  \bibinfo{journal}{Phys. Rev.} \textbf{\bibinfo{volume}{77}},
  \bibinfo{pages}{219} (\bibinfo{year}{1950}{\natexlab{a}}).

\bibitem[{\citenamefont{Yukawa}(1950{\natexlab{b}})}]{Yukawa:1950er}
\bibinfo{author}{\bibfnamefont{H.}~\bibnamefont{Yukawa}},
  \bibinfo{journal}{Phys. Rev.} \textbf{\bibinfo{volume}{80}},
  \bibinfo{pages}{1047} (\bibinfo{year}{1950}{\natexlab{b}}).

\bibitem[{\citenamefont{Efimov}(1967)}]{Efimov:1967pjn}
\bibinfo{author}{\bibfnamefont{G.~V.} \bibnamefont{Efimov}},
  \bibinfo{journal}{Commun. Math. Phys.} \textbf{\bibinfo{volume}{5}},
  \bibinfo{pages}{42} (\bibinfo{year}{1967}).

\bibitem[{\citenamefont{Alebastrov and Efimov}(1974)}]{Alebastrov:1973np}
\bibinfo{author}{\bibfnamefont{V.~A.} \bibnamefont{Alebastrov}}
  \bibnamefont{and} \bibinfo{author}{\bibfnamefont{G.~V.}
  \bibnamefont{Efimov}}, \bibinfo{journal}{Commun. Math. Phys.}
  \textbf{\bibinfo{volume}{38}}, \bibinfo{pages}{11} (\bibinfo{year}{1974}).

\bibitem[{\citenamefont{Tomboulis}(2015)}]{Tomboulis:2015gfa}
\bibinfo{author}{\bibfnamefont{E.~T.} \bibnamefont{Tomboulis}},
  \bibinfo{journal}{Phys. Rev.} \textbf{\bibinfo{volume}{D92}},
  \bibinfo{pages}{125037} (\bibinfo{year}{2015}), \eprint{1507.00981}.

\bibitem[{\citenamefont{Eliezer and
  Woodard}(1989{\natexlab{a}})}]{Eliezer:1988rr}
\bibinfo{author}{\bibfnamefont{D.~A.} \bibnamefont{Eliezer}} \bibnamefont{and}
  \bibinfo{author}{\bibfnamefont{R.~P.} \bibnamefont{Woodard}},
  \bibinfo{journal}{Phys. Rev.} \textbf{\bibinfo{volume}{D40}},
  \bibinfo{pages}{465} (\bibinfo{year}{1989}{\natexlab{a}}).

\bibitem[{\citenamefont{Eliezer and
  Woodard}(1989{\natexlab{b}})}]{Eliezer:1989cr}
\bibinfo{author}{\bibfnamefont{D.~A.} \bibnamefont{Eliezer}} \bibnamefont{and}
  \bibinfo{author}{\bibfnamefont{R.~P.} \bibnamefont{Woodard}},
  \bibinfo{journal}{Nucl. Phys.} \textbf{\bibinfo{volume}{B325}},
  \bibinfo{pages}{389} (\bibinfo{year}{1989}{\natexlab{b}}).

\bibitem[{\citenamefont{Moffat}(1990)}]{Moffat:1990jj}
\bibinfo{author}{\bibfnamefont{J.~W.} \bibnamefont{Moffat}},
  \bibinfo{journal}{Phys. Rev.} \textbf{\bibinfo{volume}{D41}},
  \bibinfo{pages}{1177} (\bibinfo{year}{1990}).

\bibitem[{\citenamefont{Evens et~al.}(1991)\citenamefont{Evens, Moffat, Kleppe,
  and Woodard}}]{Evens:1990wf}
\bibinfo{author}{\bibfnamefont{D.}~\bibnamefont{Evens}},
  \bibinfo{author}{\bibfnamefont{J.~W.} \bibnamefont{Moffat}},
  \bibinfo{author}{\bibfnamefont{G.}~\bibnamefont{Kleppe}}, \bibnamefont{and}
  \bibinfo{author}{\bibfnamefont{R.~P.} \bibnamefont{Woodard}},
  \bibinfo{journal}{Phys. Rev.} \textbf{\bibinfo{volume}{D43}},
  \bibinfo{pages}{499} (\bibinfo{year}{1991}).

\bibitem[{\citenamefont{Kleppe and Woodard}(1992)}]{Kleppe:1991rv}
\bibinfo{author}{\bibfnamefont{G.}~\bibnamefont{Kleppe}} \bibnamefont{and}
  \bibinfo{author}{\bibfnamefont{R.~P.} \bibnamefont{Woodard}},
  \bibinfo{journal}{Nucl. Phys.} \textbf{\bibinfo{volume}{B388}},
  \bibinfo{pages}{81} (\bibinfo{year}{1992}), \eprint{hep-th/9203016}.

\bibitem[{\citenamefont{Wetterich}(1993)}]{Wetterich:1992yh}
\bibinfo{author}{\bibfnamefont{C.}~\bibnamefont{Wetterich}},
  \bibinfo{journal}{Phys.Lett.} \textbf{\bibinfo{volume}{B301}},
  \bibinfo{pages}{90} (\bibinfo{year}{1993}).

\bibitem[{\citenamefont{Berges et~al.}(2002)\citenamefont{Berges, Tetradis, and
  Wetterich}}]{Berges:2000ew}
\bibinfo{author}{\bibfnamefont{J.}~\bibnamefont{Berges}},
  \bibinfo{author}{\bibfnamefont{N.}~\bibnamefont{Tetradis}}, \bibnamefont{and}
  \bibinfo{author}{\bibfnamefont{C.}~\bibnamefont{Wetterich}},
  \bibinfo{journal}{Phys.Rept.} \textbf{\bibinfo{volume}{363}},
  \bibinfo{pages}{223} (\bibinfo{year}{2002}), \eprint{hep-ph/0005122}.

\bibitem[{\citenamefont{Polonyi}(2003)}]{Polonyi:2001se}
\bibinfo{author}{\bibfnamefont{J.}~\bibnamefont{Polonyi}},
  \bibinfo{journal}{Central Eur.J.Phys.} \textbf{\bibinfo{volume}{1}},
  \bibinfo{pages}{1} (\bibinfo{year}{2003}), \eprint{hep-th/0110026}.

\bibitem[{\citenamefont{Pawlowski}(2007)}]{Pawlowski:2005xe}
\bibinfo{author}{\bibfnamefont{J.~M.} \bibnamefont{Pawlowski}},
  \bibinfo{journal}{Annals Phys.} \textbf{\bibinfo{volume}{322}},
  \bibinfo{pages}{2831} (\bibinfo{year}{2007}), \eprint{hep-th/0512261}.

\bibitem[{\citenamefont{Gies}(2012)}]{Gies:2006wv}
\bibinfo{author}{\bibfnamefont{H.}~\bibnamefont{Gies}},
  \bibinfo{journal}{Lect.Notes Phys.} \textbf{\bibinfo{volume}{852}},
  \bibinfo{pages}{287} (\bibinfo{year}{2012}), \eprint{hep-ph/0611146}.

\bibitem[{\citenamefont{Polonyi}(2006)}]{Polonyi:2006yw}
\bibinfo{author}{\bibfnamefont{J.}~\bibnamefont{Polonyi}},
  \bibinfo{journal}{Phys. Rev.} \textbf{\bibinfo{volume}{D74}},
  \bibinfo{pages}{065014} (\bibinfo{year}{2006}), \eprint{hep-th/0605218}.

\bibitem[{\citenamefont{Nagy et~al.}(2016)\citenamefont{Nagy, Polonyi, and
  Steib}}]{Nagy:2015rva}
\bibinfo{author}{\bibfnamefont{S.}~\bibnamefont{Nagy}},
  \bibinfo{author}{\bibfnamefont{J.}~\bibnamefont{Polonyi}}, \bibnamefont{and}
  \bibinfo{author}{\bibfnamefont{I.}~\bibnamefont{Steib}},
  \bibinfo{journal}{Phys. Rev.} \textbf{\bibinfo{volume}{D93}},
  \bibinfo{pages}{025008} (\bibinfo{year}{2016}), \eprint{1508.04277}.

\bibitem[{\citenamefont{Nagy et~al.}(2018)\citenamefont{Nagy, Polonyi, and
  Steib}}]{Nagy:2018fdh}
\bibinfo{author}{\bibfnamefont{S.}~\bibnamefont{Nagy}},
  \bibinfo{author}{\bibfnamefont{J.}~\bibnamefont{Polonyi}}, \bibnamefont{and}
  \bibinfo{author}{\bibfnamefont{I.}~\bibnamefont{Steib}},
  \bibinfo{journal}{Phys. Rev.} \textbf{\bibinfo{volume}{D97}},
  \bibinfo{pages}{085002} (\bibinfo{year}{2018}), \eprint{1801.08171}.

\bibitem[{\citenamefont{Coleman}(1975)}]{Coleman:1974bu}
\bibinfo{author}{\bibfnamefont{S.~R.} \bibnamefont{Coleman}},
  \bibinfo{journal}{Phys.Rev.} \textbf{\bibinfo{volume}{D11}},
  \bibinfo{pages}{2088} (\bibinfo{year}{1975}).

\bibitem[{\citenamefont{Amit et~al.}(1980)\citenamefont{Amit, Goldschmidt, and
  Grinstein}}]{Amit:1979ab}
\bibinfo{author}{\bibfnamefont{D.~J.} \bibnamefont{Amit}},
  \bibinfo{author}{\bibfnamefont{Y.~Y.} \bibnamefont{Goldschmidt}},
  \bibnamefont{and}
  \bibinfo{author}{\bibfnamefont{G.}~\bibnamefont{Grinstein}},
  \bibinfo{journal}{J.Phys.} \textbf{\bibinfo{volume}{A13}},
  \bibinfo{pages}{585} (\bibinfo{year}{1980}).

\bibitem[{\citenamefont{Nandori
  et~al.}(2001{\natexlab{a}})\citenamefont{Nandori, Polonyi, and
  Sailer}}]{Nandori:1999vi}
\bibinfo{author}{\bibfnamefont{I.}~\bibnamefont{Nandori}},
  \bibinfo{author}{\bibfnamefont{J.}~\bibnamefont{Polonyi}}, \bibnamefont{and}
  \bibinfo{author}{\bibfnamefont{K.}~\bibnamefont{Sailer}},
  \bibinfo{journal}{Phys.Rev.} \textbf{\bibinfo{volume}{D63}},
  \bibinfo{pages}{045022} (\bibinfo{year}{2001}{\natexlab{a}}),
  \eprint{hep-th/9910167}.

\bibitem[{\citenamefont{Nagy et~al.}(2007)\citenamefont{Nagy, Nandori, Polonyi,
  and Sailer}}]{Nagy:2006pq}
\bibinfo{author}{\bibfnamefont{S.}~\bibnamefont{Nagy}},
  \bibinfo{author}{\bibfnamefont{I.}~\bibnamefont{Nandori}},
  \bibinfo{author}{\bibfnamefont{J.}~\bibnamefont{Polonyi}}, \bibnamefont{and}
  \bibinfo{author}{\bibfnamefont{K.}~\bibnamefont{Sailer}},
  \bibinfo{journal}{Phys.Lett.} \textbf{\bibinfo{volume}{B647}},
  \bibinfo{pages}{152} (\bibinfo{year}{2007}), \eprint{hep-th/0611061}.

\bibitem[{\citenamefont{Pangon et~al.}(2010)\citenamefont{Pangon, Nagy,
  Polonyi, and Sailer}}]{Pangon:2009wk}
\bibinfo{author}{\bibfnamefont{V.}~\bibnamefont{Pangon}},
  \bibinfo{author}{\bibfnamefont{S.}~\bibnamefont{Nagy}},
  \bibinfo{author}{\bibfnamefont{J.}~\bibnamefont{Polonyi}}, \bibnamefont{and}
  \bibinfo{author}{\bibfnamefont{K.}~\bibnamefont{Sailer}},
  \bibinfo{journal}{Phys.Lett.} \textbf{\bibinfo{volume}{B694}},
  \bibinfo{pages}{89} (\bibinfo{year}{2010}), \eprint{0907.0496}.

\bibitem[{\citenamefont{Berezinskii}(1972)}]{Berezinskii:1971}
\bibinfo{author}{\bibfnamefont{V.~L.} \bibnamefont{Berezinskii}},
  \bibinfo{journal}{Sov. Phys.-JETP} \textbf{\bibinfo{volume}{34}},
  \bibinfo{pages}{610} (\bibinfo{year}{1972}).

\bibitem[{\citenamefont{Kosterlitz and Thouless}(1973)}]{Kosterlitz:1973xp}
\bibinfo{author}{\bibfnamefont{J.}~\bibnamefont{Kosterlitz}} \bibnamefont{and}
  \bibinfo{author}{\bibfnamefont{D.}~\bibnamefont{Thouless}},
  \bibinfo{journal}{J.Phys.} \textbf{\bibinfo{volume}{C6}},
  \bibinfo{pages}{1181} (\bibinfo{year}{1973}).

\bibitem[{\citenamefont{Mandelstam}(1975)}]{Mandelstam:1975hb}
\bibinfo{author}{\bibfnamefont{S.}~\bibnamefont{Mandelstam}},
  \bibinfo{journal}{Phys. Rev.} \textbf{\bibinfo{volume}{D11}},
  \bibinfo{pages}{3026} (\bibinfo{year}{1975}), \bibinfo{note}{[,138(1975)]}.

\bibitem[{\citenamefont{Nagy et~al.}(2009)\citenamefont{Nagy, Nandori, Polonyi,
  and Sailer}}]{Nagy:2009pj}
\bibinfo{author}{\bibfnamefont{S.}~\bibnamefont{Nagy}},
  \bibinfo{author}{\bibfnamefont{I.}~\bibnamefont{Nandori}},
  \bibinfo{author}{\bibfnamefont{J.}~\bibnamefont{Polonyi}}, \bibnamefont{and}
  \bibinfo{author}{\bibfnamefont{K.}~\bibnamefont{Sailer}},
  \bibinfo{journal}{Phys.Rev.Lett.} \textbf{\bibinfo{volume}{102}},
  \bibinfo{pages}{241603} (\bibinfo{year}{2009}), \eprint{0904.3689}.

\bibitem[{\citenamefont{Kehrein}(1999)}]{Kehrein:1999nx}
\bibinfo{author}{\bibfnamefont{S.}~\bibnamefont{Kehrein}},
  \bibinfo{journal}{Phys.Rev.Lett.} \textbf{\bibinfo{volume}{83}},
  \bibinfo{pages}{4914} (\bibinfo{year}{1999}), \eprint{cond-mat/9908048}.

\bibitem[{\citenamefont{Nandori
  et~al.}(2001{\natexlab{b}})\citenamefont{Nandori, Polonyi, and
  Sailer}}]{Nandori:2000rx}
\bibinfo{author}{\bibfnamefont{I.}~\bibnamefont{Nandori}},
  \bibinfo{author}{\bibfnamefont{J.}~\bibnamefont{Polonyi}}, \bibnamefont{and}
  \bibinfo{author}{\bibfnamefont{K.}~\bibnamefont{Sailer}},
  \bibinfo{journal}{Phil.Mag.} \textbf{\bibinfo{volume}{B81}},
  \bibinfo{pages}{1615} (\bibinfo{year}{2001}{\natexlab{b}}),
  \eprint{hep-th/0012208}.

\bibitem[{\citenamefont{Huang and Polonyi}(1991)}]{Huang:1990via}
\bibinfo{author}{\bibfnamefont{K.}~\bibnamefont{Huang}} \bibnamefont{and}
  \bibinfo{author}{\bibfnamefont{J.}~\bibnamefont{Polonyi}},
  \bibinfo{journal}{Int.J.Mod.Phys.} \textbf{\bibinfo{volume}{A6}},
  \bibinfo{pages}{409} (\bibinfo{year}{1991}).

\bibitem[{\citenamefont{Wegner and Houghton}(1973)}]{Wegner:1972ih}
\bibinfo{author}{\bibfnamefont{F.~J.} \bibnamefont{Wegner}} \bibnamefont{and}
  \bibinfo{author}{\bibfnamefont{A.}~\bibnamefont{Houghton}},
  \bibinfo{journal}{Phys.Rev.} \textbf{\bibinfo{volume}{A8}},
  \bibinfo{pages}{401} (\bibinfo{year}{1973}).

\bibitem[{\citenamefont{Kovacs et~al.}(2015)\citenamefont{Kovacs, Nagy, and
  Sailer}}]{Kovacs:2014fwa}
\bibinfo{author}{\bibfnamefont{J.}~\bibnamefont{Kovacs}},
  \bibinfo{author}{\bibfnamefont{S.}~\bibnamefont{Nagy}}, \bibnamefont{and}
  \bibinfo{author}{\bibfnamefont{K.}~\bibnamefont{Sailer}},
  \bibinfo{journal}{Phys. Rev.} \textbf{\bibinfo{volume}{D91}},
  \bibinfo{pages}{045029} (\bibinfo{year}{2015}), \eprint{1408.2680}.

\bibitem[{\citenamefont{Morris}(1994)}]{Morris:1993qb}
\bibinfo{author}{\bibfnamefont{T.~R.} \bibnamefont{Morris}},
  \bibinfo{journal}{Int. J. Mod. Phys.} \textbf{\bibinfo{volume}{A9}},
  \bibinfo{pages}{2411} (\bibinfo{year}{1994}), \eprint{hep-ph/9308265}.

\bibitem[{\citenamefont{Nagy et~al.}(2008)\citenamefont{Nagy, Nandori, Polonyi,
  and Sailer}}]{Nagy:2006ue}
\bibinfo{author}{\bibfnamefont{S.}~\bibnamefont{Nagy}},
  \bibinfo{author}{\bibfnamefont{I.}~\bibnamefont{Nandori}},
  \bibinfo{author}{\bibfnamefont{J.}~\bibnamefont{Polonyi}}, \bibnamefont{and}
  \bibinfo{author}{\bibfnamefont{K.}~\bibnamefont{Sailer}},
  \bibinfo{journal}{Phys.Rev.} \textbf{\bibinfo{volume}{D77}},
  \bibinfo{pages}{025026} (\bibinfo{year}{2008}), \eprint{hep-th/0611216}.

\bibitem[{\citenamefont{Nandori}(2008)}]{Nandori:2007zs}
\bibinfo{author}{\bibfnamefont{I.}~\bibnamefont{Nandori}},
  \bibinfo{journal}{Phys.Lett.} \textbf{\bibinfo{volume}{B662}},
  \bibinfo{pages}{302} (\bibinfo{year}{2008}), \eprint{0707.2745}.

\bibitem[{\citenamefont{Nandori}(2011)}]{Nandori:2010ij}
\bibinfo{author}{\bibfnamefont{I.}~\bibnamefont{Nandori}},
  \bibinfo{journal}{Phys.Rev.} \textbf{\bibinfo{volume}{D84}},
  \bibinfo{pages}{065024} (\bibinfo{year}{2011}), \eprint{1008.2934}.

\end{thebibliography}

\end{document}